**ORIGINAL PAPER**

# The Quantitative Comparative Economics: indices of similarity to economic systems

Ali Zeytoon-Nejad[1]

**Abstract**
This paper presents a novel quantitative approach for comparative economic studies, addressing limitations in current classification methods. Conventional approaches in comparative economics often rely on ad hoc and categorical classifications, leading to subjective judgments and disregarding the continuous nature of the spectrum of economic systems. These can result in subjectivity and significant information loss, particularly for countries with systems near categorical borders. To overcome these shortcomings, the present paper proposes distance-based indices for objective categorization, considering economic foundations and using hard data. Accordingly, the paper introduces institutional similarity indices—Capitalism Similarity Index (CapSI), Communism Similarity Index (ComSI), and Socialism Similarity Index (SocSI)—which reflect countries' positions along the economic system continuum. These indices adhere to mathematical rigor and are grounded in the mathematical fields of real analysis, metric spaces, and distance functions. By classifying 135 countries and creating GIS maps, the practical applicability of the proposed approach is demonstrated. Results show a high explanatory power of the introduced indices, suggesting their beneficial usage in comparative economic studies. The paper advocates for their adoption due to their objectivity and ability to capture structural and institutional nuances without subjective judgments while also considering the continuous nature of the spectrum of economic systems.

**JEL Classification** C00 · C43 · P10 · P20 · P51

✉ Ali Zeytoon-Nejad
zeytoosa@wfu.edu

1 School of Business, Wake Forest University, 1834 Wake Forest Rd, Farrell Hall, Building 60, Winston-Salem, NC 27109, USA

## 1 Introduction

Comparative economics involves the analysis and comparison of different economic systems across countries, aiming to develop an understanding of their foundations, structures, institutions, mechanisms, and outcomes. Rooted in a rich historical tradition, comparative economics explores the complexities of various economic systems, ranging from free market capitalism to centrally planned communism. By employing a multitude of theoretical frameworks, empirical methods, and cross-country and time-series datasets, researchers seek to uncover the factors driving economic performance and policy outcomes. Comparative economics not only sheds light on the differences between economic systems but also delves into the underlying institutional arrangements, historical trajectories, and socio-economic contexts that shape their trajectories. Through rigorous analysis and interdisciplinary inquiry, comparative economics contributes to our understanding of how the global economy operates, evolves, and impacts societies across different countries and within different economic systems.

Historically, the classification of economic systems in the field of comparative economics has predominantly depended on ad hoc and categorical methods. While these approaches have provided initial insights, they are fraught with limitations, primarily stemming from their subjectivity and oversimplified categorical frameworks, which may cause two problems. The first problem has to do with the fact that the conventional classification approaches are subject to subjective judgments. However, the scientific method is defined as the dispassionate and objective development and testing of theories about the functioning of the world. Objective study relies on objective measurement. Thus, one cannot objectively study what is not objectively measured. In comparative economics, the goal is to rigorously apply the scientific method to objectively scrutinize the types and performances of various economic systems. A crucial initial step in achieving this goal is the objective measurement, identification, and classification of countries according to their economic systems. The second issue stems from the reliance of the current classification methods on categorical classifications, which inherently overlook the reality that economic systems exist on a broad, continuous spectrum. This categorical approach to classifying economic systems of countries risks significant information loss in statistical analysis, particularly for those countries whose economic systems fall near the borders of discrete categories.

The core objective of this paper is to present a classification method that not only upholds the objectivity of economic system classifications but also prevents the significant risk of information loss in statistical inference. A reasonable approach to achieve this dual objective is to utilize a distance-based indexation. This paper introduces a novel quantitative method for objectively categorizing economic systems, leveraging their main economic foundations and hard data. Furthermore, this approach acknowledges the continuous nature of economic system morphologies. The paper presents a series of institutional similarity indices that accurately position each country's economic system on the economic system continuum. These indices account for various institutional foundations of economic systems that represent the true position of each country's economic system on the economic system continuum (both for the economy as a whole as well as separately for each main foundation of economic systems—including the type of economic organization, the type of ownership, and the size of government).

Accordingly, three similarity indices called Capitalism Similarity Index (CapSI), Socialism Similarity Index (SocSI), and Communism Similarity Index (ComSI) are introduced. These indices function as pivotal indicators of countries' placements along the economic system continuum. These indices are meticulously crafted to mirror the underlying economic foundations and institutional attributes of each country's economic system, drawing upon the mathematical fields of real analysis, metric spaces, distance functions, and functional analysis to ensure their mathematical rigor, validity, and accuracy.

In order to illustrate the practical application of the proposed methodology, a comprehensive classification of 135 countries is conducted using the aforementioned indices. Employing Geographic Information System (GIS) mapping techniques, the classifications derived from these objective similarity indices are visually depicted, offering insights into the global distribution and clustering of economic systems. The findings underscore the substantial explanatory power of the approach, underscoring its effectiveness in elucidating the structural and institutional characteristics of diverse economic systems. It is recommended that comparative economic studies investigating the effect of institutional and morphological characteristics of economic systems on different facets of economic performance should use these similarity indices, which are not only highly explanatory of differing structural and institutional forms of economic systems, but they also are objective in their constructions and are free of subjective and normative judgements.

Based on these results, there is a strong case for advocating the adoption of these similarity indices in comparative economic analyses. By surpassing the limitations inherent in traditional categorical frameworks and providing a nuanced comprehension of economic systems, this methodology shows promise in enhancing the rigor, validity, and accuracy of comparative economic analyses. In fact, the method introduced in this paper aims to serve as a major contribution to the objectivization of comparative economics. Through this research endeavor, the aim is to contribute to the advancement of comparative economics by presenting a rigorous, objective, and empirically grounded framework for evaluating the similarities and differences among economic systems, thereby enriching policy-making and scholarly discourse in the field of comparative economics.

As a testament to the robust explanatory capabilities of the similarity indices introduced in this paper, it is notable that key nations align with prevailing expert perceptions within the field. For instance, Hong Kong, Switzerland, the USA, Japan, and Australia are objectively identified among the top 20 capitalist countries by the CapSI index. Similarly, China, Russia, Venezuela, Syria, and Iran feature among the top 20 communist countries according to the ComSI index. Furthermore, the Scandinavian SocSI index identifies all five Scandinavian countries—Denmark, Finland, Iceland, Norway, and Sweden—as top 20 socialist nations, reflecting their adherence to the Nordic model.

The subsequent sections of the paper are structured as follows: The following section provides a brief literature review on the classification of economic systems in the field of comparative economics. Section 3 presents formal definitions for the three economic systems examined in this study. Subsequently, it delves into the main foundations of economic systems. Section 4 elaborates on the quantification of these foundations and provides the formulation of the similarity indices. In Section 5, empirical classifications and corresponding GIS maps are presented, analyzed, and discussed. Section 6 provides the policy implications and applications of the classification approach introduced in this paper. Section 7 draws conclusions from the overarching discussion, summarizes the key findings, and recommends areas for future research. Lastly, the paper concludes with several appendices, offering more explanations of the data based on which the empirical classifications have been carried out (Appendices 1, 2, 3).

## 2 Literature review

The proposed approach in this paper builds upon and enhances previous research in the field of comparative economics, particularly in the classification of economic systems. Prior literature has offered various typologies based on indicators such as welfare expenditure, taxation systems, and social security programs. For instance, Esping-Andersen (1990) introduces a typology that categorizes countries into liberal, conservative, and social-democratic systems based on indicators related to pension coverage and unemployment insurance. Similarly, Castles and Mitchell (1992) develop a typology that classifies countries into liberal, conservative, non-right hegemony, and radical systems, considering welfare expenditure and taxation systems. While these typologies have been valuable in understanding the diversity of economic systems, they often rely on qualitative assessments, subjective considerations, and limited economic indicators, providing only a partial view of economic systems.

Bambra (2007) critiques the over-reliance on Esping-Andersen's typology in public health research, highlighting that despite its widespread use, it has faced extensive criticism. Bambra suggests that integrating a broader range of typologies could provide a more accurate reflection in future research. Dariusz (2015) criticizes existing classifications for their failure to capture the unique features of transition economies, proposing and calling for new methods of classification. These critiques and perspectives underscore the necessity for a more nuanced classification framework, which is both theoretically grounded and empirically applicable in classifying countries into different economic systems.

Rosser and Rosser (2018) provide a comprehensive overview of comparative economic systems, emphasizing the need to accommodate the diversity of institutional arrangements within and across countries. This aligns with Pryor's (2005) argument for defining economic systems as clusters of complementary institutions rather than isolated attributes, highlighting the importance of capturing institutional interplay to reflect economic systems' true complexities. The present study builds on these foundational works by proposing a novel similarity index approach that attempts to objectively measure institutional similarities and acknowledge the continuous nature of economic system spectra.

Additionally, recent advancements in the measurement of economic systems emphasize integrating multidimensional data to reflect the hybrid nature of many modern economies. Traditional dichotomous classifications or qualitative classifications often fail to capture the nuanced transitions between economic systems, particularly in emerging markets and transitioning economies, as they gradually and morphologically change in terms of their economic institutions and structures. For example, Henisz (2000) highlights the importance of institutional constraints in shaping economic outcomes, which necessitates a framework that accounts for both economic and institutional diversity. The indices introduced in the present paper align with these advancements by offering a flexible yet precise method to quantify the degree of similarity between countries and standard economic models, thus providing a comprehensive lens for understanding economic system morphology.

Gwartney and Stroup (2014) provides a foundational overview of comparative economic systems, focusing on private property and market-directed economies, while also detailing the widespread presence of mixed economies across various countries. Chase-Dunn (1980) argues that socialist states, while representing the logic of socialism, have been shaped within and by the capitalist world economy. Zimbalist and Sherman (2014) offer a political-economic approach to comparing economic systems, emphasizing the impracticality of pure capitalism and pure socialism, and suggesting that real-world economies display degrees of both. These studies suggest the necessity for classification methods that accurately capture the degrees of similarity of these inherently mixed systems to the polar cases or standard forms of economic systems.

As discussed above, the classification of economic systems has been explored through diverse approaches in the literature. Esping-Andersen (1990) and Castles and Mitchell (1992) developed welfare-centric typologies focusing on social policies and expenditure patterns, providing insights into liberal, conservative, and social-democratic regimes. Bambra (2007) critiqued these frameworks for their narrow focus, advocating for broader and more adaptable classifications. Dariusz (2015) and Rosser and Rosser (2018) emphasized the need to account for the dynamic and hybrid nature of economic systems, particularly in transition economies. Pryor (2005) and Henisz (2000) highlighted the role of institutional clustering and political factors in shaping economic structures, aligning with Gwartney and Stroup's (1980) empirical work on economic freedom indices. Meanwhile, Chase-Dunn (1980) and Zimbalist and Sherman (2014) examined the interplay between political economy and mixed systems, challenging the notion of pure capitalism or socialism. Positioned within this landscape, this paper introduces a novel quantitative framework that leverages similarity indices to classify economic systems continuously, addressing the limitations of these categorical, static, and often subjective methodologies.[1]

---

[1] Appendix 2 provides a comprehensive table that systematically tabulates all the reviewed studies on economic system classification and summarizes the contributions and limitations of key studies and also positions the QCE framework as an innovative approach that addresses gaps in previous methodologies.

The classification of countries' economic systems into capitalist, socialist, and communist is a somewhat complex task. The studies reviewed above collectively underscore the need for a nuanced and flexible approach to classifying economic systems in such a way that the nuanced differences in the structure, characteristics, and morphology of economic systems are captured across countries quantitatively, objectively, and continuously without loss of objectivity or information. The proposed methodology in this paper is in fact an attempt to represent a significant advancement by offering a quantitative and rigorous framework for classifying countries' economic systems and assessing their similarity to major economic systems by leveraging a comprehensive set of economic foundations and mathematical tools in order to contribute to the field of comparative economics.

In light of this, the proposed methodology in this paper introduces a quantitative framework that utilizes a comprehensive set of main economic foundations, including the organization of economic activity, ownership of means of production, and extent of government intervention extracted from objective sources of hard data. This allows for a more rigorous and objective assessment of the similarity of countries' economic systems to standard forms of economic systems. The proposed framework also enables a nuanced analysis of the degree of similarity between countries' economic systems and major economic models, including capitalism, socialism, and communism. By quantifying this similarity degree, the study offers insights into the positioning of countries relative to major standard economic models and allows for the examination of intermediate or mixed economic systems without loss of important information.

## 3 Economic systems: definitions and foundations

An economic system functions as a mechanism for organizing the allocation of inputs (i.e., factors of production), the production of outputs (i.e., goods and services), and the distribution of outputs within an economy. As Nobel Laureate Milton Friedman (1974) contends, the method of economic organization and the type of economic system constitute fundamental forces that significantly shape the economic development of a country. Comprising diverse components, institutions, and entities, an economic system exerts substantial influence on the decision-making process concerning the pivotal questions confronting any economy, which in turn influences individuals' incentives and the economic performance of the economy as a whole.

The literature on the typology and morphology of economic systems, their institutions, and the impact of their institutions on their economic outcomes is replete with seminal contributions that underscore the pivotal role of economic systems in shaping economic behavior and performance. Nobel Laureate Douglas North's seminal work emphasizes institutions as fundamental determinants of economic trajectories, underscoring their role as the "rules of the game" within a society (North 1990). Building upon this foundation, subsequent studies, such as those by Acemoglu and Robinson and Johnson, delve into the empirical significance of foundations and institutions in driving economic development (Acemoglu et al. 2001; Acemoglu et al. 2005). Furthermore,

Pryor's research advocates for defining economic systems based on clusters of complementary institutions, rather than individual ones, highlighting the importance of accounting for all main institutional foundations (Pryor 2005). While existing literature explores various aspects of institutional quality and economic performance, the present paper aims to contribute uniquely by examining the institutional foundations that underpin standard types of major economic systems and introducing similarity indices that classify countries based on their economic settings objectively and with no loss of information required for accurate statistical inference. By addressing this gap and employing a comprehensive dataset spanning a wide range of countries having different economic systems, this study seeks to objectify the classification of economic systems and enhance our understanding of the intricate differences between economic systems and their main foundations. Specifically, the paper endeavors to compare and contrast the typology of three major economic systems—capitalism, socialism, and communism—and offer insights into their main institutional foundations.[2]

This paper employs the conceptual framework, definitions, and main foundations of economic systems outlined by Zeytoon-Nejad (2025). He identifies three primary foundations of economic systems: (1) the organization of economic activity, (2) the ownership of the means of production, and (3) the extent of government intervention in the economy. Drawing upon a synthesis of definitions from leading dictionaries and mainstream economic literature, Zeytoon-Nejad (2025) offers comprehensive definitions of three economic systems, which serve as the formal underpinnings of analysis in the present study. These definitions are crafted to encapsulate common themes across various sources. Accordingly, he provides the following definitions of the three main economic systems[3]:

*Capitalism* is a natural, incentive-compatible economic system to organize economic activity, which relies heavily on private ownership and free markets to organize economic activity efficiently, and resorts minimally to government intervention to deal with market failures and to play necessary public roles.

*Socialism* is a type of a mixed economic system to organize economic activity, which relies heavily on a mix of private–public ownership and adjusted markets to organize economic activity both efficiently and equally (mainly through transfer payments) and depends considerably upon government intervention to guarantee a wide range of fundamental rights, a comprehensive social safety net and welfare plans, a wide range of public goods, and a large transfer of income/wealth, which all necessitate a large tax burden to be imposed on the citizens.[4]

---

[2] While some studies may choose to explore political and legal institutions as well, the current paper exclusively concentrates on economic institutions.

[3] In the classification methodology and identification strategy that Zeytoon-Nejad (2025) used, one could only compare the two endpoints of the spectrum of economic system (i.e., capitalism and communism) while there was no possibility of classifying and making comparison of economic performance of intermediate (mixed) socialist economic systems such as the Scandinavian socialism. The present paper is also to fill this important gap in the literature.

[4] As Zeytoon-Nejad (2025) puts it, "these include a wide range of fundamental rights (a minimum standard of living, healthcare, etc.), a rather comprehensive social safety net and a wide range of welfare plans (social security, medical insurance, unemployment insurance, etc.), a wide range of public goods (public

*Communism* is a fabricated, planned economic system to organize economic activity, which relies heavily on the central premise of altruism, state ownership, and central planning to organize economic activity with a primary emphasis on equality of outcome, and is constructed fundamentally on extreme levels of government intervention in the economy, mainly through the state ownership of the means of production to ensure equality in the first place.[5]

In general, these are three major economic systems that have been extensively tested by various countries over the past several decades. In a nutshell, capitalism is characterized by free markets, private ownership, and minimal government intervention; socialism features markets with adjustments, a blend of private and public ownership, and moderate government intervention; and communism is distinguished by central planning, state ownership, and extensive government intervention. This constitutes a spectrum of economic systems, with capitalism and communism serving as the two polar extremes. The Fraser Institute publishes an annual Economic Freedom Index for numerous countries, which can serve as a numerical measure for categorizing economic systems. As Edward Lazear (2020) explains, the Fraser's index of economic freedom compiles various measures including variables such as private ownership, limited government, low taxation, and property rights, which align very closely with the dictionary definition of capitalism.[6]

Zeytoon-Nejad (2025) utilized this index, graded on a ten-point scale, to assess the economic performance of capitalism and communism, the two endpoints of the economic system continuum. Higher index values indicate proximity to laissez-faire capitalism, while lower values lean towards communism.[7] However, this indexing approach does not accommodate the study of intermediate economic systems, such as the Scandinavian form of socialism, which lies between the two extremes in terms

---

Footnote 4 (continued)

transportation, public education, etc.), a large transfer of income/wealth, which all necessitate and require a large tax burden to be imposed on the citizens and collected by the government. Socialism attempts to promote equality mainly through transfer payments after the process of production in the economy.".

[5] Communism seeks to achieve equality of outcome primarily by having the state own the means of production before the production process starts, disregarding any possible inefficiencies that may arise from state ownership.

[6] Berggren (2003) explains that the Economic Freedom Index is a significant attempt to quantify economic freedom and is reported annually in "Economic Freedom of the World" by the Fraser Institute. An alternative index is offered by the Heritage Foundation. Scholars like Berggren (2003), Mcafee (2009), and Lazear (2020) have noted the similarity between these indices in their overall implications and performance. However, the Fraser's Economic Freedom Index is favored in academic studies due to its longer data coverage since 1970, compared to Heritage's data starting only from 1995. Additionally, the Fraser's index relies on objective variables sourced from the International Monetary Fund (IMF), World Bank (WB), and World Economic Forum (WEF), while Heritage's index incorporates subjective surveys and value judgments for some sub-variables. As a result, the Fraser's index is preferred for its objectivity, making it the choice for the current study. For a detailed comparison of these indices and additional alternatives, McAfee (2009) provides comprehensive insights.

[7] Berggren (2003) defines economic freedom as a composite index aimed at characterizing the degree to which an economy is a market economy. The Fraser's Economic Freedom Index (EFI) quantifies economic freedom using a ten-point scale, with higher values indicating proximity to laissez-faire capitalism and lower values suggesting proximity to communism within the economic system of each respective country.

of the way economic organization is structured, the type of ownership of production factors, and the extent of government intervention.[8] To address this limitation, the present paper offers a parsimonious, objective, and information-retaining classification method for classifying economic systems, which is built upon a series of distance-based similarity indices. This method will allow for the examination of the morphology of intermediate cases of economic systems. The next section attends to the quantification of the aforementioned foundations and the formulization of the described similarity indices.

## 4 Economic systems: quantification of foundations and similarity indices

This section delves into the quantitative examination of the main foundations of economic systems and the development of similarity indices that can serve as indicators revealing the extent to which a country's economic system aligns with each standard economic system. It aims to provide a comprehensive, workable framework for the classification and comparison of countries' economic systems by constructing objective metrics for assessing their similarities to and differences from standard economic systems. By quantifying key dimensions such as the organization mechanisms of economic activity, ownership structures, and the size of government intervention, this section lays the groundwork for a rigorous, systematic analysis of various economic systems. At the heart of the proposed methodology lies the utilization of distance-based indices to quantify the similarities between countries' economic systems and standard economic systems. Unlike categorical classifications, which classify economic systems into discrete categories, the proposed approach here acknowledges the continuous nature of economic system morphology, thereby capturing the subtle gradations that exist within the economic system morphologies. By adopting a distance-based index approach, this study aims to mitigate the inherent subjectivity and loss of information associated with categorical classifications and thereby offer a more structurally objective and informationally comprehensive framework for analyzing economic systems.

---

[8] It is important to highlight that these three major institutions play a pivotal role in determining the typology of economic systems. These institutions are fundamental as they closely interact with incentives, which are essential drivers of economic behavior. While there are certainly other institutions that can influence economic performance, the ones mentioned above hold particular significance as they heavily shape incentives. This emphasis on these factors is not unique to this paper. Renowned Nobel Laureate in economics Milton Friedman also underscores the importance of these three institutional foundations in a joint work with her wife (Friedman and Friedman, 1999), noting that "To judge from the climate of opinion, we have won the war of ideas. Everyone—left or right—talks about the virtues of markets, private property, competition, and limited government." Others such as Rosser and Rosser (2018) in their book entitled "Comparative Economics in a Transforming World Economy" have pointed out to the great importance of these main foundations in their discussion of how one should compare economies. The approach taken here to do so offers a parsimonious index, but still allowing for the addition of more foundations if deemed necessary for specific research inquiries, thus providing flexibility without sacrificing clarity and focus.

To gain a broad perspective on the foundations of the economics system, we can consider the main institutional foundations of the three mentioned economic systems, which have to do with the ways in which they are designed to deal with (1) economic organization (ranging from free markets to central planning), (2) the ownership of means of production (ranging from private ownership to state ownership), and (3) the scope of government intervention in the economy (ranging from minimal intervention to extreme intervention). The Fraser Institute provides ten sub-indices integral to constructing their Economic Freedom Index, three of which can serve as suitable proxies for the main institutional foundations mentioned above. Among these sub-indices are free market, private ownership,[9] and small government, each accompanied by its own set of sub-indices sourced primarily from reputable institutions such as the IMF, the WB, the WEF, and other credible data sources. All the data utilized in constructing this index are derived from empirical, objective sources rather than subjective human surveys, rendering them objective measures.[10] All three indices are evaluated on a ten-point scale, where a score of ten indicates the highest degree of the respective foundation, such as private ownership. Conversely, a score of zero reflects the lowest level of that foundation.

Accordingly, if the economic system of a given country closely aligns with pure laissez-faire capitalism (located on the right end of the economic system spectrum), its main foundations will approach a score of 10 for each aspect. Conversely, as the country's main foundations move away from the capitalism endpoint of the spectrum, its economic system indicates lesser similarity to the capitalist system, resulting in lower scores for its foundations. This establishes a quantitative framework for constructing similarity indices across different economic systems. Figure 1 provides visual representations of two example countries' economic systems (depicted in blue), with the country on the left (collectively) closer to the foundations of capitalism (shown in green) and the country on the right (collectively) further away from capitalism as an economic system.[11]

Figure 2 illustrates the economic systems of two example countries (both depicted in blue) and their proximity to pure communism depicted in red (first

---

[9] In the Fraser Institute's report, there is an index named "state ownership," which is assigned a numerical value between 0 and 10. According to their explanation, a higher value on this index indicates a lower share of state or public ownership within the respective economy. To avoid potential confusion from the original name and its quantification, this index is referred to as "private ownership" in the present study. Higher values on this revised index correspond to a greater share of private ownership in the respective economy.

[10] A significant advantage of the Fraser dataset compared to other sources of economic freedom data is its transparency. The Fraser Institute not only provides the index values but also makes the underlying data used to compute each sub-index available, offering greater clarity and accessibility for researchers.

[11] As illustrated in Fig. 1, even though the country on the left has relatively less private ownership in its economy than the country on the right, it still overall resembles a standard capitalist system more closely than the country on the right. This is because the total deviation from the three main foundations of laissez-faire capitalism (represented as the overall distance of a+ b+ c) is smaller than that of the country on the right (represented as the overall distance of a'+ b'+ c'). Therefore, the left-hand-side country is more capitalist overall, despite being less capitalist in terms of only private ownership. However, to make conclusive arguments about the shape of economic systems, the overall composite scores must be calculated, and classifications must be based on the overall composite scores.

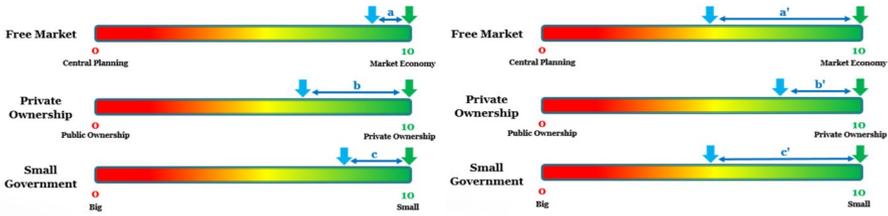

**Fig. 1** Spatial distribution of the main foundations of the economic systems of two example countries in terms of the main three institutional foundations

country) and to a specific mixed type of economic system (second country, with designated example positions of 8.5, 6.6, and 4.5 in yellow on the spectra of the degree of free market, private ownership, and government smallness).

Now, a distance-based, aggregated measure of these three individual scores of the main foundations can logically and effectively represent the overall dissimilarity of a country's economic system to the standard form of its respective economic system. These distances are computed as deviations from the foundational elements of each respective economic system. Hence, when distances are measured from, say, the benchmark positions of capitalism on the spectra, the resulting aggregated measure will indicate the degree of dissimilarity to capitalism.

To formalize this quantitative approach and establish its mathematical foundations, the mathematical fields of real analysis and functional analysis can be utilized. The former will help lay out a rigorous ground for the essence of distance, which is pivotal in defining the degrees of dissimilarity and similarity to each individual underlying foundation. The latter will help us lay out the mathematical underpinnings of aggregating the three individual distances in order to create a composite indexation of similarity. For example, a common, reasonable way to aggregate the three individual distances into one composite index is to use a summative functional form. To make the conceptual mathematical framework of this procedure more understandable, the mathematical formulation of such a summative approach for the three foundations discussed in this paper is presented first. Afterwards, the mathematical underpinnings of aggregating the three individual distances into a composite indexation of dissimilarity will be discussed using the terminologies and mathematical tools of real analysis and functional analysis.

To quantify the degree of similarity between a country's economic system and a major standard economic system, such as capitalism, this paper proposes a method based on distances from the three main foundations: the market organization (MO) of economic activity, private ownership (PO), and small government (SG). Let $D_{MO}$, $D_{PO}$, and $D_{SG}$ represent the distances of the country's economic system from the idealized position of capitalism along each of these three foundations, respectively. These distances are measured on a scale of 0 to 10, where $D_i = 0$ represents complete similarity and $D_i = 10$ represents complete dissimilarity to pure capitalism. The aggregated distance $D_A$ from capitalism can then be computed as the sum of these individual distances:

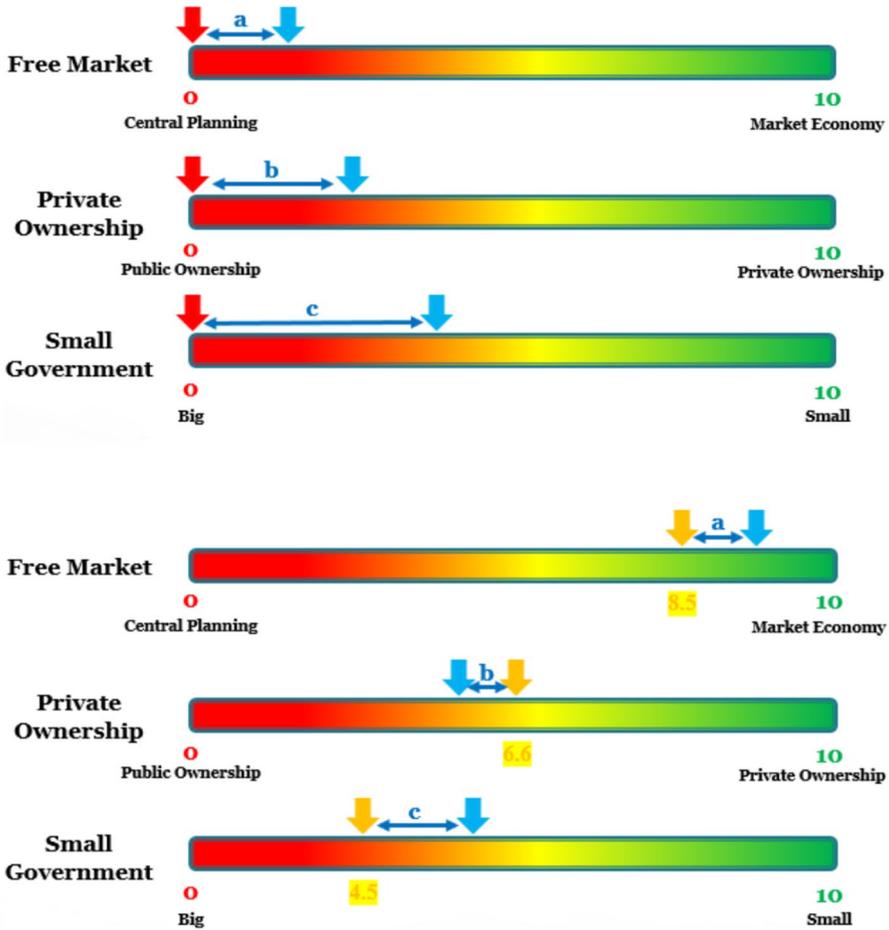

**Fig. 2** Spatial distribution of the main foundations of the economic systems of two example countries to pure communism and a specific type of a mixed economic system

$$D_A = D_{MO} + D_{PO} + D_{SG} \qquad (1)$$

Given that each individual distance ranges from 0 to 10, the maximum possible value of $D_A$ is 30, indicating the maximum dissimilarity from capitalism. Conversely, the minimum possible value of $D_A$ is 0, representing complete similarity to capitalism. To standardize and normalize the aggregated distance $D_A$ into a dissimilarity index ranging from 0 to 1, we can divide it by the maximum possible distance. Additionally, to properly express this relationship as a similarity index rather than a dissimilarity index, we need to introduce an adjustment to the formula. In fact, we need to define similarity as one minus the normalized distance from capitalism. This adjustment ensures that a country with economic foundations identical to capitalism receives a similarity index of CapSI = 1, while a country completely dissimilar from capitalism receives a similarity index of CapSI = 0. The resulting formula will be as follows:

$$\begin{aligned} \text{CapSI} &= 1 - \left\{ \frac{\sum_{i=MO,PO,SG} D_i}{\sum_{i=MO,PO,SG} MaxD_i} \right\} \\ &= 1 - \frac{(10 - S_{MO}) + (10 - S_{PO}) + (10 - S_{SG})}{30} \\ &= 1 - \frac{D_{MO} + D_{PO} + D_{SG}}{30} \end{aligned} \quad (2)$$

where CapSI denotes the Capitalism Similarity Index, $S_{MO}$ denotes the country's score on the market organization spectrum, $S_{PO}$ denotes the country's score on the private ownership spectrum, and $S_{SG}$ denotes the country's score on the smallness-of-government spectrum. As noted earlier, this yields a similarity measure between 0 and 1, where 0 indicates complete dissimilarity to capitalism and 1 indicates complete similarity to capitalism. Similar procedures can be made based on distances to create the Communist Similarity Index (ComSI) and Socialist Similarity Index (SocSI). This method provides a systematic way to quantify the similarity degree between a country's economic system and a standard economic system like capitalism, based on distances from three institutional foundations of it. It normalizes and adjusts these distances into a similarity index ranging from 0 to 1, facilitating comparative analyses and classification of economic systems.

This similarity index offers a proper framework for quantifying the degree of resemblance between economic systems. The tools of real analysis and functional analysis can be leveraged to provide and ensure a foundation of mathematical rigor for this similarity indexation method. At the core of this framework lies the concept of normed spaces, which serve as the mathematical underpinning for measuring distances between economic systems. Let $X$ be the normed vector space of economic systems, equipped with a norm $\| \cdot \|$ that quantifies the "distance" between two economic systems (say, one being a country's economic system and another being a standard economic system). This normed vector space allows for the measurement of distances between economic systems using the norm function, which assigns a non-negative, real number to each vector in the space, satisfying certain properties such as positivity, non-degeneracy, scalability, and the triangle inequality, ensuring that the introduced distance measures adhere to mathematical norms and principles.

In the context of comparative economics, economic systems can be represented as vectors in $X$, with each component of the vector corresponding to an institutional foundation of the economic system (e.g., the type of organizing economic activity, type of ownership, and size of government intervention). Let $x$ and $y$ denote two such economic systems, represented as vectors in $X$. The distance between $x$ and $y$ can be computed using norm-based distance functions, such as the Manhattan distance based on the L1 norm:

$$d(x,y) = \sum_{i=1}^{3} |x_i - y_i| \text{ with the L1 norm as } \|x\|_1 = \sum_{i=1}^{n} |x_i| \quad (3)$$

This distance function encapsulates the dissimilarity between economic systems, allowing for precise measurement and comparison of the similarity of a country's economic system to a standard, benchmark economic system. Furthermore, the proposed similarity index capitalizes on functional analysis techniques to aggregate the distances between economic systems across multiple dimensions. Let $d_{MO}$, $d_{PO}$, and $d_{SG}$ denote the distances between a country's economic system, $x$, and a benchmark economic system, $y$, along the economic activity organization, ownership, and government intervention dimensions, respectively. These distances can be combined, normalized, and adjusted for similarity, as shown below, to obtain an aggregated similarity index as follows:

$$SI(x,y) = 1 - \left\{ \frac{1}{3}\left( \frac{d_{MO}}{\max(d_{MO})} + \frac{d_{PO}}{\max(d_{PO})} + \frac{d_{SG}}{\max(d_{SG})} \right) \right\} \qquad (4)$$

where $SI(x,y)$ denotes the similarity index between the two economic systems under study, and $\max(d_{MO})$, $\max(d_{PO})$, and $\max(d_{SG})$ represent the maximum possible distances along each dimension, ensuring that the dissimilarity measure falls within the range of 0 to 1, and the normalized distance index of dissimilarity is subtracted from 1 in order to convert into its corresponding similarity index. Through the application of real analysis and functional analysis techniques and methodology, the similarity indices proposed in this paper are ensured to offer a mathematically rigorous and computationally reliable approach to comparing economic systems. By leveraging mathematical objects, concepts, and tools such as norms, distance functions, vectors, and normed spaces, the proposed methodology in this paper establishes a strong, reliable foundation for conducting objective quantitative analyses in the realm of comparative economics, specifically in the classification of economic systems across various countries.

Computing the above-mentioned similarity index for the two polar cases of economic systems is relatively straightforward. For CapSI, it involves computing $CapSI = 1 - \frac{(10-S_{MO})+(10-S_{PO})+(10-S_{SG})}{30}$ as the sum of distances (dissimilarities) from 10 and adjusted for similarity, whereas for ComSI, it involves computing $ComSI = 1 - \frac{(S_{MO}-0)+(S_{PO}-0)+(S_{SG}-0)}{30}$ as the sum of distances (dissimilarities) from 0 and adjusted for similarity. However, the computation of this SI for an intermediate, mixed socialist country presents more complexity. In what follows, this procedure is shown and explained for the case of the Scandinavian form of socialism, commonly known as the Nordic model, as defined in the previous section.

The Scandinavian form of socialism encompasses the economic norms prevalent in the Nordic countries, including Denmark, Finland, Iceland, Norway, and Sweden. Key components entail a regulated, market-based economy with a blend of private and state ownership, alongside an extensive welfare state supported by substantial taxation and government intervention. Given the high degree of similarity among these countries in terms of their economic systems and underlying foundations, we can delineate a standardized form of Scandinavian socialism by computing the average scores of the foundations of their economic systems. Consequently, these countries will cluster around these averages, and any deviations observed in other countries' scores from these averages will signify their divergence from the standard Scandinavian socialist economic system.[12] Then, $SocSI = 1 - \frac{|S_{MO}-S^*_{MO}|+|S_{PO}-S^*_{PO}|+|S_{SG}-S^*_{SG}|}{30}$ where $S^*_{MO}$, $S^*_{PO}$, and $S^*_{SG}$ are the Scandinavian socialism benchmarks derived from the Nordic countries' average score values.

**Table 1** The mean scores of the economic system foundations across the five Nordic countries

| Number | Scandinavian countries | Free Market Index | Private Ownership Index | Small Government Index |
|---|---|---|---|---|
| 1 | **Denmark** | 8.39 | 7.01 | 4.77 |
| 2 | **Finland** | 7.69 | 6.92 | 5.17 |
| 3 | **Iceland** | 7.79 | 7.65 | 6.35 |
| 4 | **Norway** | 7.50 | 6.80 | 5.43 |
| 5 | **Sweden** | 7.78 | 7.94 | 4.67 |
|  | **Average** | **7.83** | **7.26** | **5.28** |

**Table 2** The Socialism Similarity Index (SocSI) for each of the Nordic nations along with the collective average score for the Scandinavian SocSI

| Number | Scandinavian countries | Socialist Similarity Index, SocSI (7.83, 7.26, 5.28) |
|---|---|---|
| 1 | **Denmark** | 0.95 |
| 2 | **Finland** | 0.98 |
| 3 | **Iceland** | 0.93 |
| 4 | **Norway** | 0.96 |
| 5 | **Sweden** | 0.94 |
| - | **Average** | **0.95** |

When employing average calibration to objectively identify Nordic economies, all Nordic economies are identified as such using the proposed SocSI method. This will be shown in the next section. The specifics of this calibration are depicted in two tables. Table 1 displays the mean scores of the main economic foundations across the five Nordic countries. Table 2 illustrates the SocSI for each of these nations, alongside the collective average score, which serves as the benchmark SocSI for the prevalent Scandinavian socialist economic system observed in these five countries.

Arranging the scores for free market, private ownership, and small government in a sequential numerical order within parentheses, the standard form of capitalism can be represented as CapSI (10,10,10), communism as ComSI (0,0,0), and the average scores of the Nordic countries' main economic system foundations yield SocSI (7.83,7.26,5.28). For brevity, these are abbreviated in this paper as standard CapSI, ComSI, and SocSI, respectively. The following section presents world maps illustrating the CapSI, ComSI, and SocSI for the 135 countries in the dataset, indicating the degrees of similarity to capitalism, communism, and socialism, respectively, for each country's economic system (Figs. 3, 4, 5, 6, 7, 8).

---

[12] It is important to note that this logic and methodology can be applied to any other intermediate, mixed economic system characterized by a distinct combination of the three main foundations without loss of generality.

## 5 Economic systems: classifications and map representations

This section provides the visual representation of economic system classifications across 135 countries using the Communism Similarity Index (ComSI), Capitalism Similarity Index (CapSI), and Socialism Similarity Index (SocSI). These maps serve as useful tools to illustrate the degrees of similarity that each country's economic system bears to communism, capitalism, and socialism, respectively. By providing a comprehensive overview of economic structures on a global scale, these visualizations offer invaluable insights into the distribution and clustering of economic systems across different regions and continents. This categorization discerns patterns and trends that shed light on the prevalence and influence of different economic systems. In the forthcoming series of maps, the 135 countries present in the dataset are depicted with their corresponding similarity scores to the three specified standard economic systems. Following each world map, the top 20 countries in terms of each similarity index are reported for further discussion of the results of the proposed classification methodology and for examining the effectiveness of the introduced similarity indices. This will help in offering insights into the study of the alignments of the results of the proposed classification method in this paper with those provided by prior classifications in the existing literature.

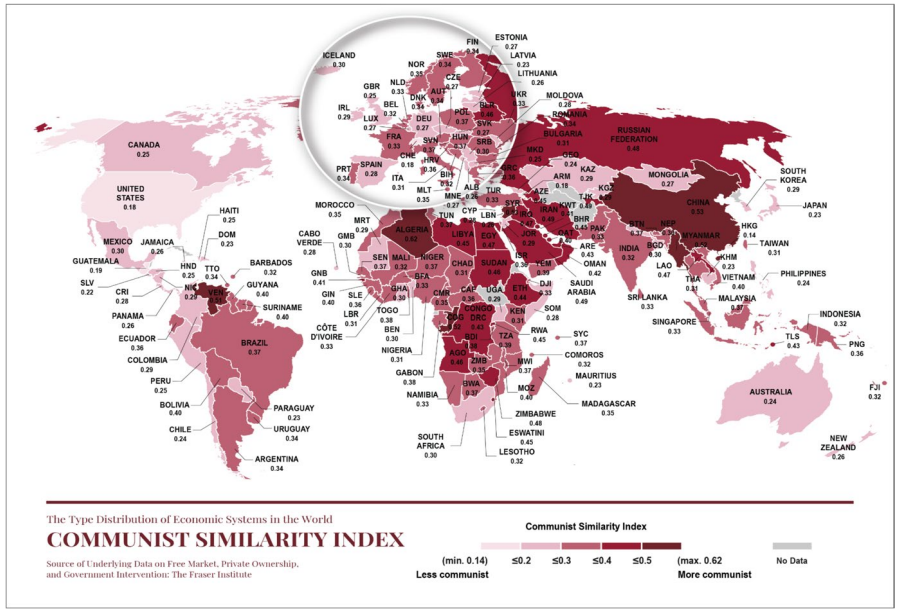

**Fig. 3** Average Communist Similarity Index (ComSI) for each country over the period 1995–2020

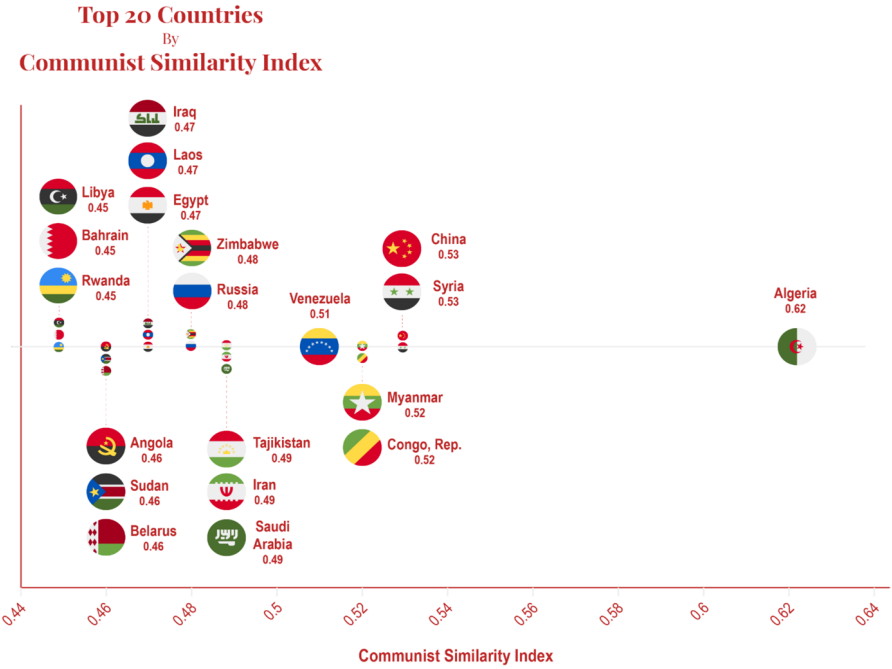

**Fig. 4** Top 20 countries by Communist Similarity Index (ComSI) over the period 1995–2020

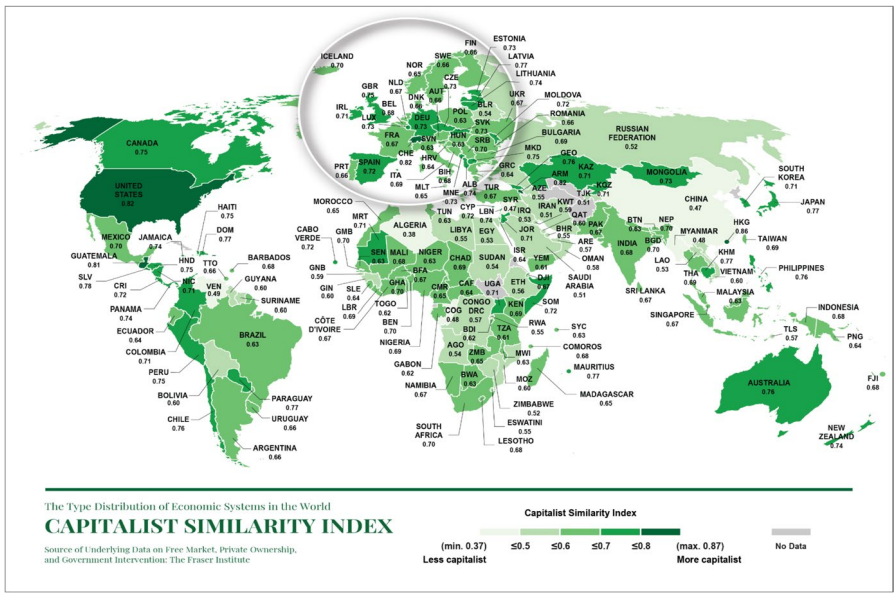

**Fig. 5** Average Capitalist Similarity Index (CapSI) for each country over the period 1995–2020

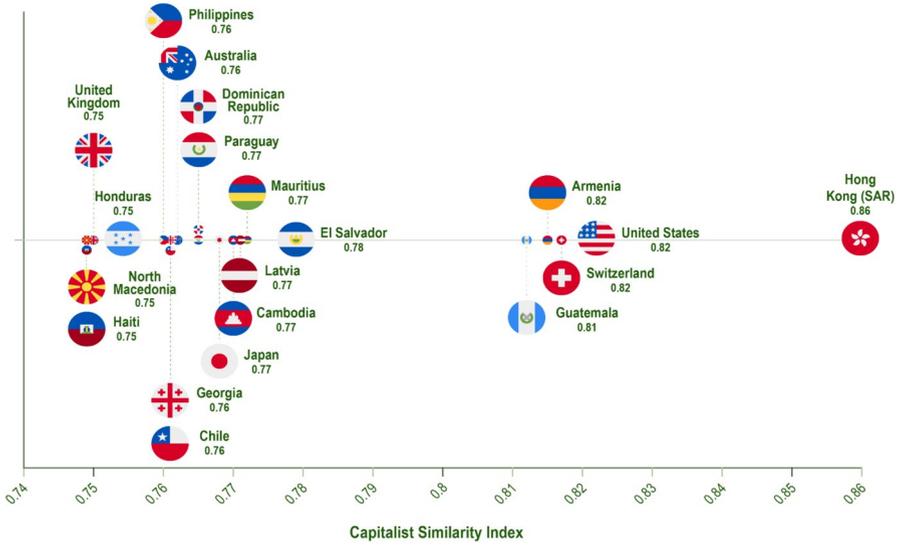

**Fig. 6** Top 20 countries by Capitalist Similarity Index (CapSI) over the period 1995–2020

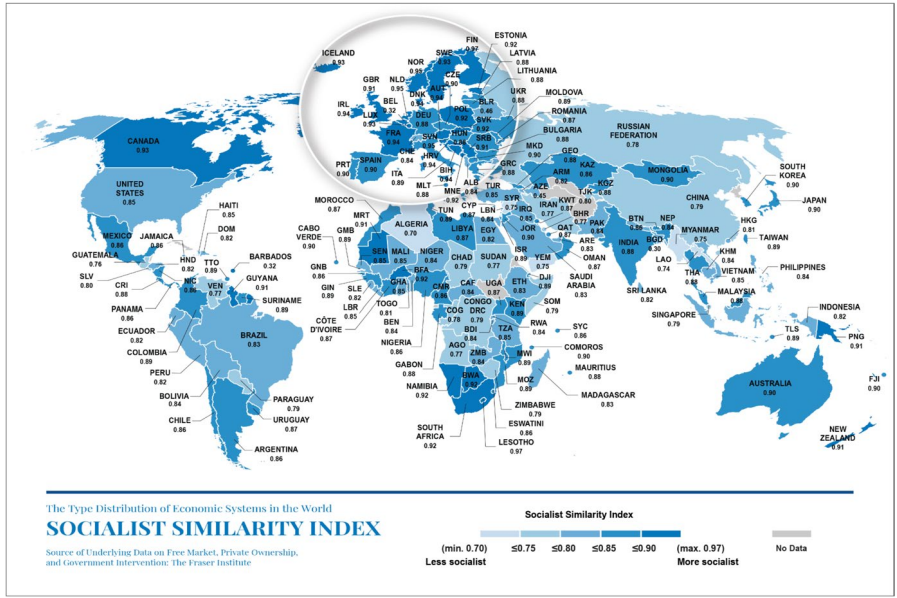

**Fig. 7** Average Socialist Similarity Index (SocSI) for each country over the period 1995–2020

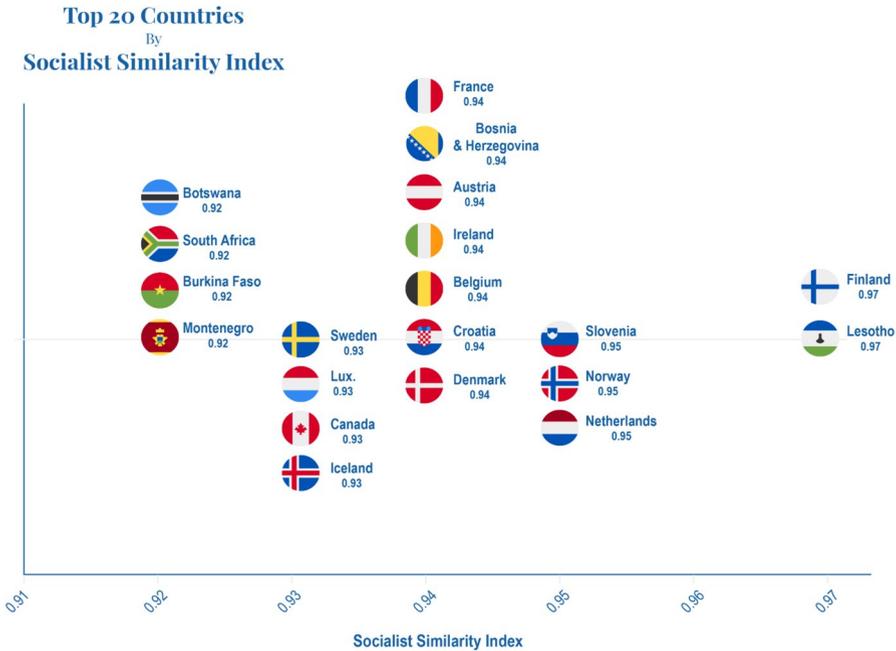

**Fig. 8** Top 20 countries by Socialist Similarity Index (SocSI) over the period 1995–2020

The results reported above affirm the high explanatory power of the methodology proposed in this paper for objectively classifying countries under various economic systems. Notably, the results align closely with the mainstream consensus among subjective classifications from prominent sources such as Esping-Andersen (1990), Bornstein (1994), Gregory and Stuart (2004), Ferragina and Seeleib-Kaiser (2011), and Gregory and Stuart (2014). However, the method proposed here accomplishes this classification task objectively using hard data and with reference to the main foundations of economic systems. For instance, the CapSI index effectively identifies Hong Kong, Switzerland, the USA, Japan, and Australia among the top 20 capitalist countries, consistent with common subjective perceptions among experts and scholars in the field.[13] Similarly, the ComSI index accurately identifies China,

---

[13] There are some countries (such as Armenia) whose economic system in our analysis here—which considers the past three decades—have been identified as being similar to the main foundations of capitalism by the methodology used in this paper, while these countries do not generally exhibit high levels of per capita income or economic development as Western countries (such as the USA) do. It is crucial to make a clear distinction between "the type of economic system" that a country adopts (which has to do with similarity of its economic system to the main foundations of a standard economic system during the study period of this paper) and "the level of economic development" of the country (which has mainly to do with which stage of economic development the country has been over the course of the study period in this paper, which is measured usually by factors such as GDP per capita, industrialization, development of financial markets, and infrastructure). While some countries may exhibit similarities in the morphology of their economic systems, their levels of economic development can vary significantly. This underscores the need to differentiate between the "type of economic system" a country adopts and its overall "level of economic prosperity and development." For example, a country may adhere closely to the principles and foundations of capitalism, socialism, or communism, yet its economic-development

Russia, Venezuela, Syria, and Iran among the top 20 communist countries. Remarkably, the Scandinavian SocSI identifies all five Scandinavian nations—Denmark, Finland, Iceland, Norway, and Sweden—among the top 20 socialist nations, reflecting their adherence to the Nordic model.

Moreover, echoing the characterization by Ferragina and Seeleib-Kaiser (2011), several countries beyond the Nordic region are significantly linked with Scandinavian socialism. These nations are termed as secondary democratic-socialist states by Ferragina and Seeleib-Kaiser (2011). Remarkably, several social-democratic nations from these additional countries, including Austria, Belgium, Canada, France, and the Netherlands, emerge among the top 20 countries resembling Scandinavian socialism. The identification of this secondary group of countries as Scandinavian socialists by the proposed methodology in this paper, due to their economic systems closely mirroring the Nordic model, within the top 20 socialist countries reinforces the credibility and validity of the classification methodology outlined in this paper.

Compared to prior classifications of countries into economic systems carried out by others, the methodology presented in this paper offers several notable advantages. First of all, it prioritizes the main foundations of economic systems rather than political or social considerations. While some existing classifications have focused more on other aspects like political ideologies (e.g., Zimbalist and Sherman 2014), legal systems (e.g., Porta et al. 1998), or social structures (e.g., Fukuyama 1995), the approach presented in this paper concentrates on the core economic foundations that underpin a country's economic system and influence the behavior of economic agents on a large scale. Secondly, the similarity indices proposed in this paper offer a parsimonious model of objectively classifying countries in terms of their economic systems. This stands in contrast to highly composite statistics, such as the Economic Freedom Index computed by organizations such as the Fraser Institute or the Heritage Foundation, which can be overly aggregated and probably too

---

Footnote 13 (continued)

status may be characterized by a high or low level of GDP per capita.

  Therefore, when analyzing and comparing economic systems, it is crucial to consider the distinction between the "type of system" and the "level of economic development" to gain a comprehensive understanding of a country's economic landscape. Thus, it is natural to observe that not all the countries whose form of economic system (in terms of the main foundations of their economic systems) mirror closely those of developed, capitalist countries have high levels of economic development as well. In fact, Western countries which started embracing foundations of capitalism during 1800s started growing and developing sooner and faster than other countries that started later to do so, and this time lag resulted in an economic phenomenon called the Great Divergence. The presence of such non-developed countries in the top 20 capitalist countries implies that their foundations of economic system in the past two or three decades have mirrored those of developed, capitalist countries, but it is important to recognize that this CapSI scoring contains no information about the level of economic development of those non-developed or developing countries, which does not have much to do with the current morphology of their economic systems, so their level of economic development is a separate consideration. In other words, economic development is a result of continuing to operate on a well-functioning type of economic system for centuries and not solely a year or a decade, and the record of history has shown that, in general, the nations that chose to operate within market-based systems have achieved higher levels of economic performance and development historically. For more information, see Zeytoon-Nejad (2025).

complex. The proposed methodology in this paper simplifies the classification process while retaining its effectiveness.

Thirdly, the methodology and similarity indices introduced here are sufficiently comprehensive to be adaptable to various forms of economic systems, extending beyond the two polar cases of capitalism and communism and also covering intermediate cases such as Scandinavian socialism, as exercised in this paper. Unlike other indices put forth in the literature of comparative economics, such as the Fraser's Economic Freedom Index, which can accommodate only the two polar cases, and as such is limited to studying only specific economic systems on the two endpoints of the economic system spectrum, the introduced approach here allows for defining any standardized forms of intermediate systems on the spectrum. This versatility enables researchers to study the performance of a wide range of economic systems without sacrificing generality and applicability.

Fourthly, the introduced similarity indices in this paper are objective and computed through a mathematically grounded methodology based on hard data, rather than relying on subjective opinions. This stands in stark contrast to the Economic Freedom Index of the Heritage Foundation, which utilizes survey-based data and constructs its indices based on some subjective judgments. By leveraging factual data rather than subjective assessments, the methodology proposed in this paper provides a more objective and dependable classification framework.

Fifthly, the methodology proposed in this paper offers quantitative measures of similarity to standard economic systems, rather than solely relying on qualitative, categorical classifications, which can result in a loss of information. Instead, the proposed approach preserves useful information that can be utilized for statistical inference when analyzing the characteristics and performance of these economic systems. Given the continuous nature of the spectrum of economic systems and their main foundations, this quantitative framework proves essential, especially considering the fact that nearly all countries exhibit some degree of being a mixed economic system.

Sixthly, unlike other approaches, this methodology does not confine the classification of each country to only a single economic system. Instead, it allows for the computation of a degree of similarity to each type of economic system for every country in consideration. This is a notable advantage of the method introduced in this paper. Take, for instance, the US economic system, which boasts a CapSI score of 0.82, ranking it second among the most capitalist countries globally and indicating its strong alignment with laissez-faire capitalism. At the same time, despite not being among the top 20 Scandinavian socialist countries, its similarity index to Scandinavian socialism stands at 0.85, shedding light on the extent to which the US economic system has adopted aspects of a welfare state in recent decades. Lastly, with a ComSI score of 0.18, the US economic system ranks among the least communist countries worldwide, providing a comprehensive overview of its economic system.

It is worth emphasizing that this paper maintains a neutral stance (with respect to the relative importance of the main foundations in the overall construct of the economic system) by assigning equal weights to the three main foundations of economic systems in its indexation process. This approach tries to avoid potential subjectivity associated with determining weights for each foundation. However, should there be a compelling argument favoring the emphasis of one foundation over others in defining economic system typology, adjustments to the formulation of similarity indices can be easily made. Furthermore, it is important to recognize that the classification model outlined in this paper (covering only three main foundations of economic systems) can be regarded simply as the baseline model for the Quantitative Comparative Economics (QCE) approach.[14] This baseline model offers scholars a starting point, allowing for modifications such as adding or altering foundations and adjusting the importance weights to suit specific research purposes. Through this flexibility, researchers can tailor the model to best fit their analytical needs and enhance the applicability of QCE in diverse economic studies.

All in all, this paper offers compelling evidence of how the QCE approach can effectively classify countries into different economic systems based on factual, hard data without losing useful information that can be used in statistical inference to improve the reliability of the results of statistical tests. By prioritizing positivity, dispassionateness, and impartiality in the classification process, this approach contributes to the objectification and advancement of economic system classifications in the social sciences community. The consistency of results generated by this method with the mainstream of prior classifications underscores its effectiveness in objectively identifying countries' economic systems through the similarity indices introduced in this paper and using the foundation indices that are all based on objective, hard data. Therefore, the categorization of countries using the QCE approach is objective and free of subjectivity. This methodological rigor not only enhances the credibility of economic system classifications but also lays the groundwork and paves the way for objective comparative analyses. Researchers can leverage the QCE framework to explore the impact of the choice of economic systems on various indicators of socio-economic outcomes with greater accuracy and objectivity.

---

[14] As noted earlier, these three dimensions were selected for a multitude of reasons. First, they represent the fundamental institutional foundations that define the structure and functionality of any economic system. Market organization captures how economic activities are coordinated, ranging from free markets to central planning, which directly impacts resource allocation, efficiency, and innovation. Ownership reflects the distribution of control over production resources, from private ownership to state ownership, influencing incentives, property rights, and wealth creation. Government size represents the extent of government intervention in the economy, encompassing taxation and public spending, which shape the scale of public goods provision and redistribution. These dimensions are the common themes in most definitions of economic systems and are also widely recognized in the literature as critical for distinguishing between economic systems. For instance, works by Friedman (1974) and North (1990) emphasize market coordination and institutional quality, while Pryor (2005) and Acemoglu and Robinson (2012) discuss ownership structures and government intervention as key determinants of economic outcomes. Moreover, the Fraser Institute's Economic Freedom Index, a widely used measure in comparative economics, includes sub-indices aligned with these pillars, further underscoring their salience. By focusing on these three pillars, the QCE approach provides a parsimonious yet comprehensive framework for classifying economic systems. While additional contextual factors such as cultural or political dimensions can complement these foundations, market organization, ownership, and government size offer a universally applicable and empirically measurable foundation for distinguishing economic systems.

## 6 Policy implications and applications

The Quantitative Comparative Economics (QCE) framework offers a powerful tool for policymakers, researchers, and international organizations seeking to assess, compare, and reform economic systems. By providing objective, data-driven indices that measure the degree of similarity of national economies to capitalism, socialism, and communism, QCE enables more informed decision-making in various policy contexts. This section explores the specific applications of the QCE methodology in international comparisons, economic reform evaluations, and investment decisions.

### 6.1 Applications in international comparisons

QCE indices provide a systematic means of comparing economic systems across countries, offering a more nuanced alternative to binary or categorical classifications. This is particularly valuable for organizations such as the World Bank, International Monetary Fund (IMF), and Organisation for Economic Co-operation and Development (OECD), which frequently analyze cross-country economic performance, institutional structures, and policy outcomes. By applying QCE indices, these institutions can more accurately assess the economic policy environment in different nations and track how institutional changes influence long-term trajectories of economic growth.

For example, comparisons between emerging economies and advanced economies using QCE indices can reveal whether economic transitions are leading to more market-oriented systems or increased government intervention. Policymakers in transition economies, such as Vietnam, China, or post-Soviet states, can use these indices to benchmark their economies against different economic models and strategically plan policy adjustments based on their intended economic direction.

### 6.2 Economic institutions, economic reform evaluations, and policy adjustments

Since economic policies operate within institutional frameworks, their effectiveness depends on the underlying economic system. Policies that perform well in one institutional setting may yield different results in another due to variations in market structures, government intervention, and property rights. The QCE methodology provides a systematic way to analyze these interactions, allowing researchers and policymakers to examine how shifts in institutions influence economic performance. By assessing how changes in regulations, ownership structures, or government policies affect economic outcomes, QCE enables a more context-sensitive evaluation of policy success. Furthermore, it helps test the sensitivity of economic performance to institutional changes, offering insights into the stability and adaptability of policy measures across different economic system conditions. This makes QCE a powerful tool for understanding the institutional determinants of policy success and designing tailored economic strategies.

Governments undertaking structural economic reforms can use QCE indices to evaluate the effectiveness of their policies over time. By tracking changes in a country's similarity to capitalism, socialism, or communism, policymakers can assess whether institutional transformations align with their intended objectives. This approach is particularly relevant for countries undergoing major economic shifts, such as market liberalization (e.g., China's economic reforms since the 1980s), post-socialist transitions (e.g., Eastern Europe after the fall of the Soviet Union), or state-driven economic models like Singapore's hybrid system. Additionally, reform-driven governments can benchmark their policies against successful economies by comparing their Capitalism Similarity Index (CapSI) with historically market-driven countries, like Switzerland and the USA, or their Socialism Similarity Index (SocSI) with Scandinavian welfare states to evaluate their alignment with the Nordic model.

### 6.3 Investment decisions and business strategy

The QCE methodology can also guide investment decisions by multinational corporations (MNCs), sovereign wealth funds, and institutional investors. Foreign investors often evaluate the economic policy environment of potential investment destinations, considering factors such as market openness, regulatory structures, and government intervention. QCE indices provide quantifiable measures of institutional similarity, helping investors identify economies that align with their risk tolerance and strategic interests.

For example, a firm specializing in technology and innovation may prefer economies with high CapSI scores, indicating market-driven institutions and strong private-sector protections. In contrast, a company in the public infrastructure or healthcare sectors may prioritize economies with moderate SocSI scores, where governments play an active role in allocating more resources to such industries. By offering an objective measure of economic system alignment, QCE indices help businesses navigate international markets, assess regulatory risks, and develop country-specific strategies.

Overall, the QCE methodology offers policymakers, international organizations, and investors a robust, data-driven framework for analyzing and comparing economic systems. By applying QCE indices, decision-makers can track institutional transformations against benchmark economies and assess the effectiveness of economic reforms. However, given the complexity of economic systems, QCE should be used alongside other qualitative and quantitative analyses to ensure well-rounded policy decisions. The adaptability of the QCE approach allows for future refinements, including the incorporation of additional economic dimensions, making it a useful, flexible tool for guiding economic policy worldwide.

## 7 Conclusion and summary

This paper presented a novel approach termed Quantitative Comparative Economics (QCE) for objectively classifying countries based on their economic systems. Central to this methodology are the institutional similarity indices—Communism

Similarity Index (ComSI), Socialism Similarity Index (SocSI), and Capitalism Similarity Index (CapSI)—which serve as key indicators of countries' positions along the economic system continuum. These indices are meticulously designed to reflect the underlying economic foundations and institutional characteristics of each country's economic system parsimoniously and draw upon principles from the mathematical sub-branch of real analysis, metric spaces, distance functions, and functional analysis to ensure their mathematical rigor, reliability, and validity.

The proposed approach builds upon and enhances previous research in the field of comparative economics, particularly in the classification of economic systems. Previous typologies have primarily focused on qualitative distinctions between economic systems, such as liberal, conservative, and social-democratic. In contrast, the proposed framework in this paper enables a nuanced analysis of the degree of similarity between countries' economic systems and major standard economic systems, including capitalism, socialism, and communism. By quantifying this similarity degree, the study offers insights into the positioning of countries relative to standard economic systems and also allows for the examination of intermediate or mixed economic systems. By leveraging a set of economic foundations and mathematical tools, the study provides a more nuanced understanding of the typology and classification of economic systems, contributing to the field of comparative economics.

The results of the study demonstrate the effectiveness and reliability of the QCE approach in objectively identifying countries' economic systems, where the CapSI index identifies Hong Kong, Switzerland, the USA, Japan, and Australia among the top 20 capitalist countries, while China, Russia, Venezuela, Syria, and Iran are reported among the top 20 communist countries, and Denmark, Finland, Iceland, Norway, and Sweden are identified among the top 20 socialist nations.[15] Furthermore, the study highlights the importance of distinguishing between the "type of economic system" adopted by a country and its "level of economic development." While some countries may exhibit similarities in their economic systems, their levels of economic prosperity and development can vary significantly. It is crucial to consider this distinction when analyzing and comparing economic systems to gain a comprehensive understanding of each country's overall economic setting.

The QCE approach offers several notable advantages over prior methods of classifying economic systems. It focuses on the main foundations of economic systems and provides a parsimonious model that is adequately comprehensive and flexibly adaptable to various forms of economic systems. Importantly, the QCE methodology relies on objective data rather than subjective judgments, enhancing the reliability of the classification framework. It allows for a nuanced understanding by accommodating multiple economic systems for each country and provides quantitative measures of similarity to every standard economic system, facilitating detailed comparative analyses. Overall, the QCE approach represents a significant advancement in economic system classification and contributes to the objectification of economic system classification within the social sciences community.

---

[15] It is important to note that no predefined benchmark classification has been imposed in this study; rather, the visualization of results was based solely on the top 20 scores for each similarity index. While these benchmarks serve as a theoretically and empirically grounded starting point, there could be value in exploring data-driven methods, such as clustering analysis, to identify groupings of economic systems. These methods could reveal emergent patterns and further validate or refine the benchmarks used in the QCE framework. Future research could employ clustering techniques such as k-means clustering or hierarchical clustering to identify natural groupings of economic systems based on the same underlying data, the same similarity indices, and the same scoring procedures introduced by the QCE methodology. Integrating such clustering methods into the QCE framework would enhance the empirical grounding of benchmark positions and, in particular, provide further insights into the nuanced structures of mixed or hybrid economies. In fact, such incorporation can enrich the QCE framework and its empirical grounding, as well as its applications in comparative economic studies.

In conclusion, the QCE approach presented in this study provides a useful and flexible framework for researchers and policy analysts active in the field of comparative economics. By providing an objective and quantitative classification framework, QCE facilitates more accurate analyses of the impact of the morphology of economic systems on various aspects of economic performance. For instance, the similarity indices introduced in this paper open up new avenues for research into the impact of economic system typology on various indicators of economic performance. Moreover, the QCE method facilitates the examination of temporal trends in economic system morphology, as well as the assessment of dynamic changes in the form and output of transitioning economies. Moving forward, the author encourages further research and refinement of the QCE approach to continue advancing the baseline QCE method introduced in this paper and increase our understanding of economic system classifications.

For future research purposes, researchers can attend to methodological refinements, continuously refining the QCE methodology by incorporating new variables,[16] improving measurement techniques, and enhancing mathematical and statistical modeling approaches. They can also attempt to undertake a cross-national validation, validating the QCE approach across a broader range of countries and regions[17] and over different time horizons to assess its applicability and robustness in diverse socio-economic settings over different time horizons.[18] They can also conduct longitudinal studies using the QCE approach to track changes in economic system classifications over time and analyze their implications for economic development.[19] Furthermore, sub-national analyses can be conducted using QCE, extending the application of the QCE approach to analyze economic system classifications at sub-national levels, such as states or provinces, to uncover regional disparities and policy implications. Moreover, the QCE approach can be integrated with qualitative methods, exploring the integration of qualitative research methods with the QCE approach to provide, say, a more comprehensive understanding of the socio-political dynamics shaping economic systems. These avenues of research have the potential to culminate in the formulation of evidence-based policy recommendations derived from QCE findings. These recommendations can serve as vital inputs for decision-makers, promoting economic policy formulation, economic development, and ultimately social welfare.

---

[16] While the present QCE approach, which focuses on market organization, ownership, and government size as the core pillars, should be thought of as the baseline or canonical version of QCE, alternative frameworks can include additional dimensions that could complement or extend the analysis. Incorporating additional contextual dimensions related to, say, welfare provision, income redistribution (as emphasized by Esping-Andersen 1990), labor market arrangements, corporate governance, financial systems (as emphasized by Hall and Soskice 2001), governance quality, democratic accountability, political stability (as emphasized by North, 2009), societal trust, and social capital (as emphasized by Fukuyama 1995) could provide deeper insights into economic systems and more effectively capture institutional nuances depending on what the main research questions of interest are.

---

[17] It is also important to recognize the potential for the Modifiable Areal Unit Problem (MAUP) when applying the QCE framework at different geographic scales. While the current study classifies economic systems at the national level, economic policies and institutions may vary significantly at subnational levels. Future studies could extend the QCE methodology to regional- or state-level analyses, evaluating whether the classifications remain consistent when applied to different spatial aggregations.

[18] It can be argued that economic systems are not isolated constructs; instead, they can be deeply embedded within the political and social structures of countries. Factors such as governance frameworks, cultural norms, and societal fabrics can then influence the organization and outcomes of economic systems significantly. Incorporating these dimensions into the QCE framework may present an avenue for future research to provide a more holistic analysis of economic systems.

[19] It is important to note that economic systems of countries are dynamic entities that evolve over time, often undergoing significant transitions in their form and typology. The QCE framework is well-suited to capture such morphological transformations, enabling researchers to trace the progression of transition economies and analyze how changes in their institutional foundations affect their economic policies and overall economic performance. This adaptability can position QCE as a powerful tool for studying the evolution of economic systems across diverse contexts.

## Appendix 1. List of countries in the dataset

See Table 3.

**Table 3** Countries in the dataset

| # | Country | # | Country | # | Country |
|---|---|---|---|---|---|
| 1 | Albania | 46 | France | 91 | Namibia |
| 2 | Algeria | 47 | Gabon | 92 | Netherlands |
| 3 | Angola | 48 | Gambia, The | 93 | New Zealand |
| 4 | Armenia | 49 | Georgia | 94 | Nicaragua |
| 5 | Australia | 50 | Germany | 95 | Niger |
| 6 | Austria | 51 | Ghana | 96 | Nigeria |
| 7 | Azerbaijan | 52 | Greece | 97 | Norway |
| 8 | Bangladesh | 53 | Guatemala | 98 | Oman |
| 9 | Barbados | 54 | Haiti | 99 | Pakistan |
| 10 | Belarus | 55 | Honduras | 100 | Panama |
| 11 | Belgium | 56 | Hong Kong | 101 | Paraguay |
| 12 | Benin | 57 | Hungary | 102 | Peru |
| 13 | Bhutan | 58 | Iceland | 103 | Philippines |
| 14 | Bolivia | 59 | India | 104 | Poland |
| 15 | Bosnia and Herzegovina | 60 | Indonesia | 105 | Portugal |
| 16 | Botswana | 61 | Iran, Islamic Rep | 106 | Romania |
| 17 | Brazil | 62 | Ireland | 107 | Russian Federation |
| 18 | Bulgaria | 63 | Israel | 108 | Rwanda |
| 19 | Burkina Faso | 64 | Italy | 109 | Saudi Arabia |
| 20 | Burundi | 65 | Jamaica | 110 | Senegal |
| 21 | Cambodia | 66 | Japan | 111 | Serbia |
| 22 | Cameroon | 67 | Jordan | 112 | Sierra Leone |
| 23 | Canada | 68 | Kazakhstan | 113 | Singapore |
| 24 | Cape Verde | 69 | Kenya | 114 | Slovak Republic |
| 25 | Chad | 70 | Korea, Rep | 115 | Slovenia |
| 26 | Chile | 71 | Lao PDR | 116 | South Africa |
| 27 | China | 72 | Latvia | 117 | Spain |
| 28 | Colombia | 73 | Lebanon | 118 | Sri Lanka |
| 29 | Congo, Dem Rep | 74 | Lesotho | 119 | Suriname |
| 30 | Congo, Rep | 75 | Liberia | 120 | Sweden |
| 31 | Costa Rica | 76 | Lithuania | 121 | Switzerland |
| 32 | Cote d'Ivoire | 77 | Luxembourg | 122 | Tajikistan |
| 33 | Croatia | 78 | Macedonia | 123 | Tanzania |
| 34 | Cyprus | 79 | Madagascar | 124 | Thailand |
| 35 | Czech Republic | 80 | Malawi | 125 | Togo |
| 36 | Denmark | 81 | Malaysia | 126 | Tunisia |
| 37 | Dominican Republic | 82 | Mali | 127 | Turkey |
| 38 | Ecuador | 83 | Malta | 128 | Uganda |

**Table 3** (continued)

| 1 | Albania | 46 | France | 91 | Namibia |
|---|---|---|---|---|---|
| 39 | Egypt, Arab Rep | 84 | Mauritius | 129 | Ukraine |
| 40 | El Salvador | 85 | Mexico | 130 | UK |
| 41 | Estonia | 86 | Moldova | 131 | USA |
| 42 | Eswatini | 87 | Mongolia | 132 | Uruguay |
| 43 | Ethiopia | 88 | Morocco | 133 | Venezuela, RB |
| 44 | Fiji | 89 | Mozambique | 134 | Zambia |
| 45 | Finland | 90 | Myanmar | 135 | Zimbabwe |

## Appendix 2. Tabulated list of studies reviewed from the literature on economic system classification frameworks

Table 4 provides a comparative summary of key studies on economic system classification, highlighting their focus, limitations, and contributions to the field. It positions the Quantitative Comparative Economics (QCE) framework within this literature, indicating its advantages in addressing gaps through a rigorous, data-driven approach, and underscores how QCE offers a more objective, flexible, and comprehensive methodology for analyzing economic systems.

This table provides a clear synthesis of the literature and highlights how the QCE framework addresses gaps and improves upon previous classification methods. By incorporating objective data, mathematical rigor, and a continuum-based analysis, QCE provides a more nuanced and adaptable methodology for understanding economic systems.[20] The advantages of QCE can be summarized as objectivity (using hard data and mathematical rigor, avoiding subjective biases), continuity (capturing economic systems along a continuum rather than in rigid categories), flexibility (adapting to hybrid, transitional, and mixed systems, accommodating real-world complexities), granularity (enabling "detailed" temporal, cross-country, and panel analyses), and replicability (providing a replicable and transparent methodology for future research). These advantages make QCE a useful tool for academic research, policy analysis, and institutional analysis.

---

[20] While the QCE methodology relies on objective data from reputable sources such as the Fraser Institute's Economic Freedom Index, it is important to acknowledge that no data source is entirely free from limitations and biases. For instance, differences in data availability, measurement errors, or variations in institutional reporting standards across countries could introduce biases to the classifications generated by QCE. To address these potential limitations, researchers should leverage robustness checks, such as sensitivity analyses, to assess the impact of missing data or measurement errors on the indices and their resulting classifications.

**Table 4** A review of economic system classifications and the positioning of Quantitative Comparative Economics (QCE) within the related literature

| Study | Focus/contribution | Limitations | Relation to QCE and QCE relative advantages |
|---|---|---|---|
| Esping-Andersen (1990) | Typology of welfare capitalism: liberal, conservative, and social-democratic systems based on social policies and welfare structures | Welfare-centric; limited focus on broader economic foundations | QCE extends beyond welfare policies to include broader economic foundations (market organization, ownership, and government size), offering a more comprehensive classification |
| Castles and Mitchell (1992) | Classification based on welfare expenditure and taxation systems, identifying liberal, conservative, and radical regimes | Narrow focus on fiscal aspects, overlooking institutional and structural diversity | QCE integrates structural and institutional attributes, providing a more nuanced and multidimensional view of the morphology of economic systems |
| Bambra (2007) | Critique of over-reliance on narrow typologies in public health and welfare research, advocating for broader frameworks | Lack of an alternative quantitative classification method | QCE addresses this gap with a mathematically grounded, quantitative approach that avoids categorical rigidity and enables continuous spectrum analysis |
| Dariusz (2015) | Highlights the inadequacies of existing frameworks in capturing transition economies, calling for innovative classification methods | Limited practical methodology for addressing dynamic and hybrid systems | QCE explicitly accommodates hybrid and transitional systems through similarity indices, ensuring adaptability and objectivity across diverse economies including hybrid, emerging, and transition economies |
| Rosser and Rosser (2018) | Emphasizes the dynamic and evolving nature of economic systems in response to globalization and institutional changes | Does not provide a quantitative methodology for capturing system transitions | QCE captures temporal changes in economic systems with its flexible framework, enabling time-series, longitudinal, cross-sectional, and panel analyses |
| Pryor (2005) | Advocates clustering complementary institutions to define economic systems, emphasizing the importance of institutional synergy | Relies on qualitative methods; lacks a replicable quantitative framework | QCE builds on institutional clustering but quantifies it with distance-based similarity indices, enhancing replicability and objectivity |

**Table 4** (continued)

| Study | Focus/contribution | Limitations | Relation to QCE and QCE relative advantages |
|---|---|---|---|
| Henisz (2000) | Explores the influence of political institutions on economic structures and performance | Focuses narrowly on political dimensions without integrating core economic foundations | QCE integrates political-economic dimensions within a broader framework, capturing the interplay of governance and economic-institutional attributes |
| Gwartney and Stroup (2014) | Develops the Economic Freedom Index, focusing on private property and market-directed economies | Overly aggregated and limited to a binary spectrum (economic freedom vs. restriction) | QCE offers a more granular analysis, accommodating mixed systems and the visibility of their institutional nuances using hard data by calculating similarity indices across a continuum |
| Chase-Dunn (1980) | Examines the influence of capitalist world systems on socialist economies, emphasizing the hybrid nature of real-world systems | Does not offer a framework to quantify hybrid systems or degrees of similarity to pure models | QCE quantitatively captures hybrid systems and provides nuanced measures of similarity to benchmark economic systems |
| Zimbalist and Sherman (2014) | Critiques the impracticality of pure capitalism or socialism, emphasizing the prevalence of mixed systems in the real world | Lacks a quantitative method for analyzing the spectrum of mixed systems | QCE provides a systematic way to classify mixed systems using objective data and mathematical rigor, bridging the gap between theoretical critique and empirical application |

# Appendix 3. Data sources and variable definitions

Table 5 details the data sources, variable definitions, and processing steps used to construct the similarity indices.

**Table 5** Data sources, variable definitions, and processing steps

| Variable | Definition | Data source | Notes on processing |
|---|---|---|---|
| Market organization (MO) | Measures the degree of market reliance in organizing economic activity | Fraser Institute's Economic Freedom Index Dataset | Sub-index "regulation" is used as a proxy to the degree to which markets are free. This sub-index is in turn computed using a multitude of other sub-indexes indicating regulation of credit, labor, and businesses, which are all accounted for by the Fraser Institute. According to their explanations, the greater the extent to which a country allows markets to determine prices and refrain from regulatory activities, the higher this index score ends up being in the respective economy |
| Private ownership (PO) | Measures the extent of private ownership of resources and production factors | Fraser Institute's Economic Freedom Index Dataset | Sub-index "state ownership," which is assigned a numerical value between 0 and 10 by the Fraser Institute, is used as a proxy for private ownership. According to their explanations, a higher value on this index indicates a lower share of the state ownership of means of production within the respective economy |
| Government size (GS) | Measures the extent of government intervention in economic activities, including taxation and spending | Fraser Institute's Economic Freedom Index Dataset | Sub-index "size of government" is directly used to quantify this dimension, which in turn is composed of several sub-indexes related to government taxes, subsidies, transfers, government investment, government consumption, and the like, and are computed and aggregated by the Fraser Institute |

Table 5 (continued)

| Variable | Definition | Data source | Notes on processing |
|---|---|---|---|
| Communism Similarity Index (ComSI) | Composite measure of alignment with characteristics of communism | Derived from above indices | Calculated as a decreasing function of distances from benchmark communism scores (0, 0, 0) for MO, PO, and GS |
| Capitalism Similarity Index (CapSI) | Composite measure of alignment with characteristics of capitalism | Derived from above indices | Calculated as a decreasing function of distances from benchmark capitalism scores (10, 10, 10) for MO, PO, and GS |
| Socialism Similarity Index (SocSI) | Composite measure of alignment with Scandinavian socialism characteristics | Derived from above indices | Benchmarked using the average MO, PO, and GS scores of the Nordic countries, being (7.83, 7.26, 5.28), and then calculated as a decreasing function of distances from these benchmarks |

Variable definitions: Each variable reflects a key institutional foundation of economic systems—market reliance, private ownership, and government size. Data source details: All data are derived from the Fraser Institute's Economic Freedom dataset, which compiles information from the World Bank, IMF, and other international organizations. Data processing: Variables were normalized to a 0–10 scale to facilitate comparison. Distances from benchmark scores were aggregated using a summative functional form, ensuring mathematical rigor and consistency. Benchmarks: Benchmark positions for capitalism (10,10,10) and communism (0,0,0) are based on theoretical ideals, while Scandinavian socialism benchmarks were empirically derived using Nordic country averages, as explained in the paper. The original source of data that support the findings of this study are openly available at https://www.fraserinstitute.org/economic-freedom/dataset

**Supplementary Information** The online version contains supplementary material available at https://doi.org/10.1007/s10368-025-00665-9.

**Funding** Open access funding provided by the Carolinas Consortium.

**Data availability** The data that support the findings of this study are openly available at https://www.fraserinstitute.org/economic-freedom/dataset. Additionally, the dataset of similarity indices introduced in this paper will be made publicly available and updated annually at https://aznejad.com/my-dashboards/.

## Declarations

**Conflict of interest** The author declares no competing interests.